# COSMOS: A CONTEXT SENSITIVE MODEL FOR DYNAMIC CONFIGURATION OF SMARTPHONES USING MULTIFACTOR ANALYSIS

K.S. Kuppusamy[1], Leena Mary Francis[2], G. Aghila[3]

[1]Department of Computer Science, School of Engineering and Technology, Pondicherry University, Pondicherry, India
[3]Department of Computer Science & Engineering, National Institute of Technology Puducherry, India
[1]`kskuppu@gmail.com`, [2]`rosebeauty02@gmail.com`, [3]`aghilaa@gmail.com`

## *ABSTRACT*

*With the prolific growth in usage of smartphones across the spectrum of people in the society it becomes mandatory to handle and configure these devices effectively to achieve optimum results from it. This paper proposes a context sensitive model termed COSMOS (COntext Sensitive MOdel for Smartphones) for configuring the smartphones using multifactor analysis with the help of decision trees. The COSMOS model proposed in this paper facilitates the configuration of various smartphone settings implicitly based on the user's current context, without interrupting the user for various inputs. The COSMOS model also proposes multiple context parameters like location, scheduler data, recent call log settings etc to decide the appropriate settings for the smartphones. The proposed model is validated by a prototype implementation in the Android platform. Various tests were conducted in the implementation and the settings relevancy metric value of 90.95% confirms the efficiency of the proposed model.*

## 1. INTRODUCTION

The mobile phones have become an integral part of day-to-day life for majority of the people in the modern society. The evolution of smartphones from the traditional mobile devices has opened up a critical research domain which attempts to provide solutions to the issues surrounding these smartphone devices. Surpassing all other traditional communication channels, smartphones are raising as a favourite choice of people, not only for voice call purposes but also for accessing the internet which includes mail access, social networking, mobile shopping and mobile banking. [1]

Though the smartphone devices are very powerful in nature, the efficient handling of these devices depends on optimal settings which have to be explicitly done by the user in majority of the scenarios. The manoeuvring of these settings requires a level of expertise with respect to these devices which cannot be assured in all occasions. The improper settings in these smartphone devices lead to the substandard performance which drastically affects the optimal utility level of these devices. Adding complexity to this scenario is the dynamicity of these optimal settings. The optimal configuration varies across the temporal dimension for a single user, based on various context settings. The optimal settings for different users do differ due to the personalization requirements.

This research attempts to provide a solution to this problem by proposing a context sensitive model termed as COSMOS (COntext Sensitive MOdel for Smartphones). The COSMOS model relieves the user of the smartphone from the task of manually configuring the device for various settings like screen brightness, vibration mode, ringtone volume, GPS and WiFi settings etc. The objectives of this research work are as listed below:

- Proposing a context sensitive model for configuration of smartphones to achieve optimal performance.
- Devising various context parameters to support the model in choosing the best suited settings based on user's current context.

The remainder of the paper is organised as follows: In section 2 the related works carried out in this area which have motivated this research work has been discussed. Section 3 deals about the proposed model and its mathematical representation. Section 4 focuses on the implementation of the model in android platform and results of the implementation work. Sections 5 conclude the research paper and highlight the future implementation of this research work.

## 2. MOTIVATIONS

Apart from being simple voice communication devices, the mobile phone has evolved into very powerful multipurpose devices with many mission critical applications. [2] As the context surrounding the mobile device is varying, applying context specificity becomes an important task. The context sensitive handling of mobile devices with respect to security has been studied in detail by researchers. [3]

The on-field usage of smartphones with respect to various users have been studied for their long term implications. [4]  The study of configuring the smartphones optimally becomes a critical research problem as they can lead to proper utilization of important resources in the smartphones like battery utilization etc. [5], [6] As the user base of the smartphones have become mammoth, the study of analysing how a large group of users handle their smartphones is also an important research issue which has been addressed by researchers. [7] It has been concluded by researchers that usage context is an important parameter with respect to efficient handling of smartphone devices. [8]

The computation of context of the smartphones is studied by researchers for inquiry and action in mobile devices. [9] Context sensitivity in mobile devices is studied for various purposes like advertising, rich user interfaces etc. [10] The proposed research work, COSMOS utilizes the context of the smartphones for efficient manoeuvring of settings in the smartphones.

For the purpose of analysing the input vectors, this research work utilizes decision trees. The decision tree falls under the machine learning classification category. [11]–[13] There are other classification techniques like Support Vector Machine , k- Nearest neighbour etc [14], [15]. The reasons for utilizing the decision tree in the proposed model are due to their ability to perform well with the discontinuous and missing data.

## 3. THE COSMOS MODEL

This research work proposes a model termed COSMOS (Context Sensitive MOdel for Smartphones). The block diagram of the COSMOS model is as shown in Figure 1. It has two major components viz., COSMOS Client and COSMOS server. The individual components of the COSMOS model are as listed below:

- **Location Tracker**: The role of location tracker component is to track the current location of the device using device's built-in GPS facility or using the WiFi connectivity. Based on the current location, the settings of the smartphone would get altered, which is done by analysing the past data and the corresponding user action with respect to the specified location. For example after reaching a meeting hall, if the user has put his / her smartphone in silent mode in the past then, based on the location the phone would be put in silent mode automatically when user reaches that place.
- **Scheduler Interface**: The role of Scheduler interface is to fetch the current schedules like "Meeting" from the device's built-in scheduler. The inputs from the scheduler provide critical data about the user's current context based on which the settings would be orchestrated.
- **Call Log Fetcher:** The role of Call Log Fetcher is to fetch the recent calls from the device's call log repository. The number of calls to be fetched is customized by a call log window length. The rationale for incorporating the call log fetcher in the model is to check the immediate previous calls the user has either made or received. Based on these calls and their past associations the settings would be chosen.

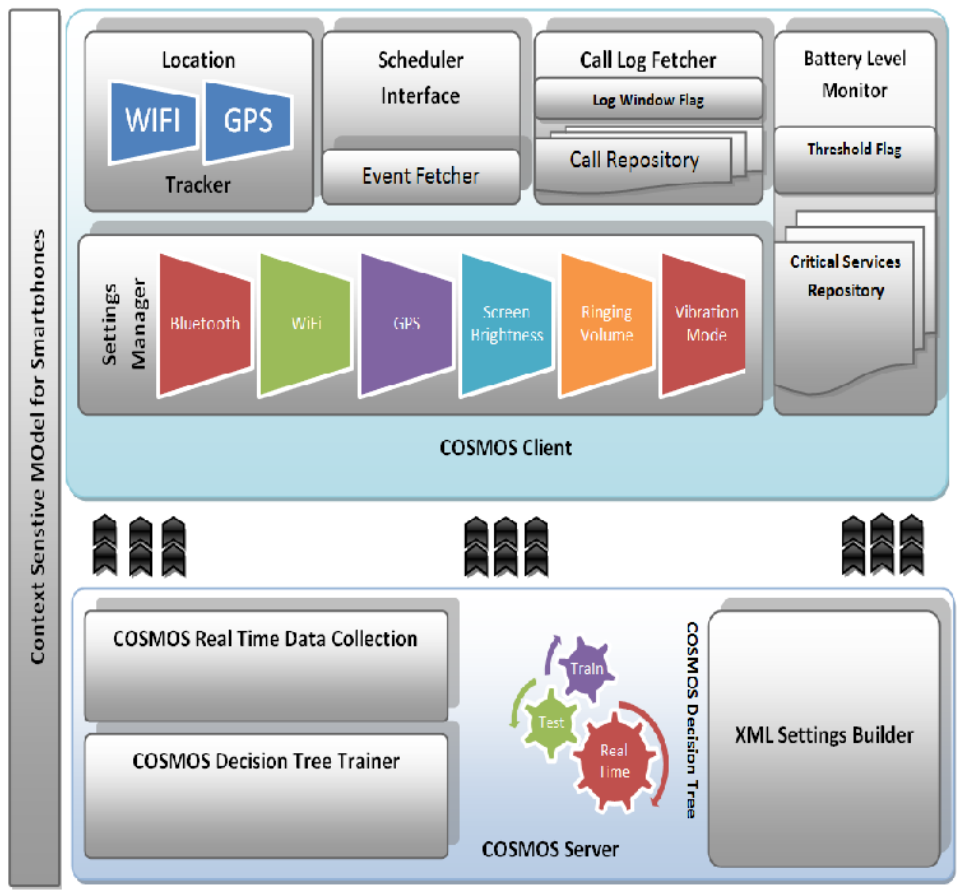

Figure 1. The COSMOS Model Block Diagram

- **Battery Level Monitor**: The role of battery level monitor is to fetch the amount of the remaining power from the device's battery. Based on the remaining battery level few settings would be adjusted. For example if the battery level goes below a critical threshold level all the power hungry components like Bluetooth, WiFi shall be set to off. The critical services repository holds the list of settings which should not be altered even during the low battery scenario.
- **Settings Manager**: The objective of Settings Manager is to load best-suited settings based on the inputs from various other components listed above. The settings manager primarily handles six different settings viz., Bluetooth, WiFi, GPS, Screen Brightness, Ring Volume and Vibration Mode.
- **COSMOS Server Component**s: The major components of the COSMOS server are "Real time data collection", "Decision Tree Trainer" , "Decision Tree" and XML settings builder. The role of real time data collection is to gather the input vector from the client. The decision tree trainer is to train the COSMOS server based on the input data received from the client. The XML settings builder would compile the target settings decided by the decision tree, which is to be sent to the client.

The context of the user is gathered with the help of various inputs received from Location Tracker, Scheduler Interface, Call Log Fetcher and Battery Level monitor. These inputs are analyzed with the help of decision trees and the corresponding the settings are chosen.

In order to reduce to power requirements in the smartphones the COSMOS model is divided into two major blocks. The client component is loaded in the device itself and the another component is loaded in the COSMOS server which would be communicated by the COSMOS clients with the inputs sensed and the decision would be made in the COSMOS server and the result would be reverted back to the client in the form of a XML file. This XML file would be interpreted by the COSMOS client to make the necessary settings changes in the device as instructed by the COSMOS server. In case, if the internet access channels like WiFi or devices data settings are disabled then the communication between the COSMOS client and server would be established with the help of Short Message Service (SMS).

## 3.1 The Mathematical Model

This section deals with the mathematical representation of the COSMOS model. The four factors utilized in the COSMOS model are represented in (1).

$$\Omega = \begin{Bmatrix} \alpha \\ \beta \\ \chi \\ \delta \end{Bmatrix} \quad (1)$$

In (1), the multifactor set is represented as $\Omega$. The Location tracker component is represented as $\alpha$, the scheduler interface is represented as $\beta$, the call log fetcher as $\chi$, the battery level monitor as $\delta$.

The Location tracker gathers the location either with the help of GPS or WiFi which is represented as shown in (2). In (2) $\alpha_g$ indicates the GPS component and $\alpha_w$ indicates the WiFi component.

$$\alpha = \begin{Bmatrix} \alpha_g \\ \alpha_w \end{Bmatrix} \quad (2)$$

The scheduler interface fetches the current schedule of the user from the device's built-in scheduler. The events related to the current time slot alone are fetched. The time slot window shall be dynamically adjusted based on the user's requirements. The $\beta$ component is represented as shown in (3).

$$\beta = \begin{Bmatrix} \forall \lambda \in \Phi & if\ (T_{NOW} - \kappa \leq T(\lambda) \leq T_{NOW} + \kappa) : \beta = \beta \cup \lambda \\ \varnothing & otherwise \end{Bmatrix} \quad (3)$$

In (3), the set of events from the scheduler is shown as $\Phi$. The time-slot window is shown as $\kappa$. If the time of the scheduler event $T(\lambda)$ falls in the current time slot window then it is appended with $\beta$. If none of the events are within the current time slot window then an empty set $\varnothing$ is returned.

The call log fetcher, which fetches the recent calls to and from the device within a predefined time slot, which is represented as shown in (4).

$$\chi = \begin{cases} \forall \mu \in \Pi & if\ (T(\mu) \geq T_{Now} - \kappa): \chi = \chi \cup \mu \\ \varnothing & otherwise \end{cases} \quad (4)$$

The set of all calls to and from the device within a predefined time slot $\kappa$ is represented as $\mu$. All these call information are added to the set. If no such calls are available in the predefined time slot then an empty set is returned as shown in (4).

For the battery power setting $\delta$, if the level goes below the critical level $\omega$, then the power crisis flag $\rho$ is set, as shown in (5).

$$\delta = \begin{cases} if\ (\delta \leq \omega): & set\ \rho\ as\ ON \\ \varnothing & otherwise \end{cases} \quad (5)$$

After computing all the four parameters, the set $\Omega$ is sent to the COSMOS server for the selection of the optimal setting.

In the COSMOS server the settings are analyzed using the decision trees. The decision trees are used to pick the values from the predefined ranges in case of continuous data. If the data is of the Boolean type like Bluetooth (On, Off) then either the true or false value is chosen. As the decision trees requires adequate training, the COSMOS model trains the decision tree based on the real time data that it receives from the COSMOS client.

$$\Delta = Train(\alpha\ \ \beta\ \ \chi\ \ \delta) \quad (6)$$

During this period the COSMOS server doesn't provide any settings suggestions. Once the decision tree is sufficiently trained, then the COSMOS server provides the optimal settings suggestions.

$$\Delta s = \{P, \Sigma, T, \Upsilon, \Psi, \Theta\} \quad (7)$$

In (7), $\Delta s$ holds six different settings. $P$ represents the Bluetooth, $\Sigma$ represents the GPS, $T$ indicates WiFi, $\Upsilon$ represents the screen brightness, $\Psi$ indicates the ringing volume and $\Theta$ represents the vibration mode.

For the COSMOS model, the decision tree algorithm utilized is J48 which is a variation of the ID3 algorithm. The reason for choosing the J48 is due to the open source nature of it and the ability to handle discontinuous factor data. Based on the input vector received from the client, the COSMOS decision tree selects any of the predefined settings which serve as the close match.

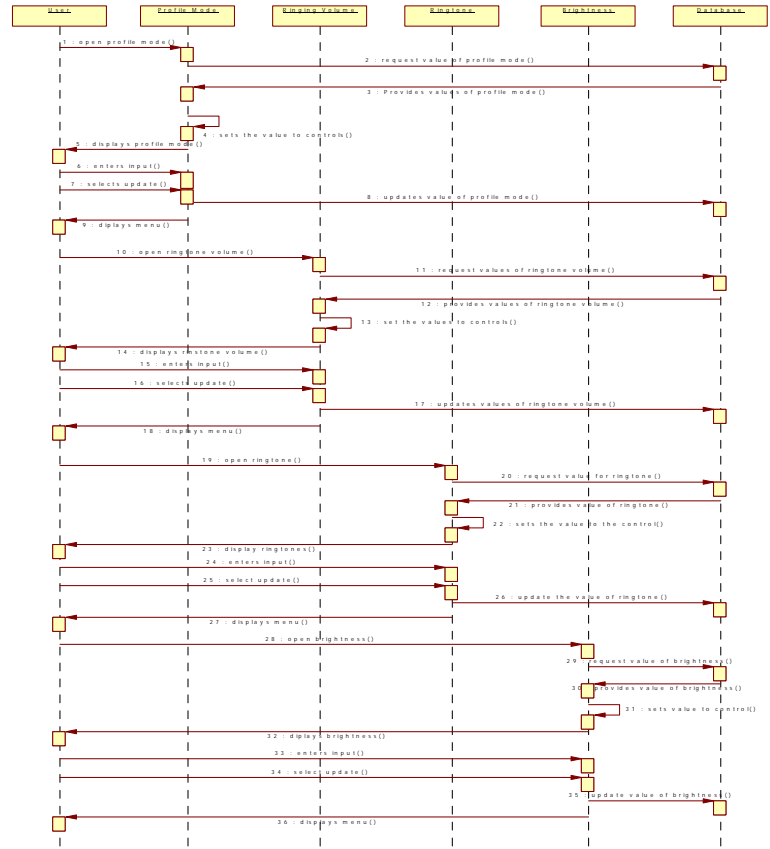

Figure 2 : Sequence Diagram – COSMOS

## 4. THE IMPLEMENTATION

The proposed context sensitive model for dynamic configuration of smartphones has been implemented in the Android 2.3 Operating System. [16] The COSMOS client is loaded into the smartphone device. The server component is implemented as a PHP based web application with Apache as the web server. The sequence diagram of the COSMOS model is illustrated in Figure 2.

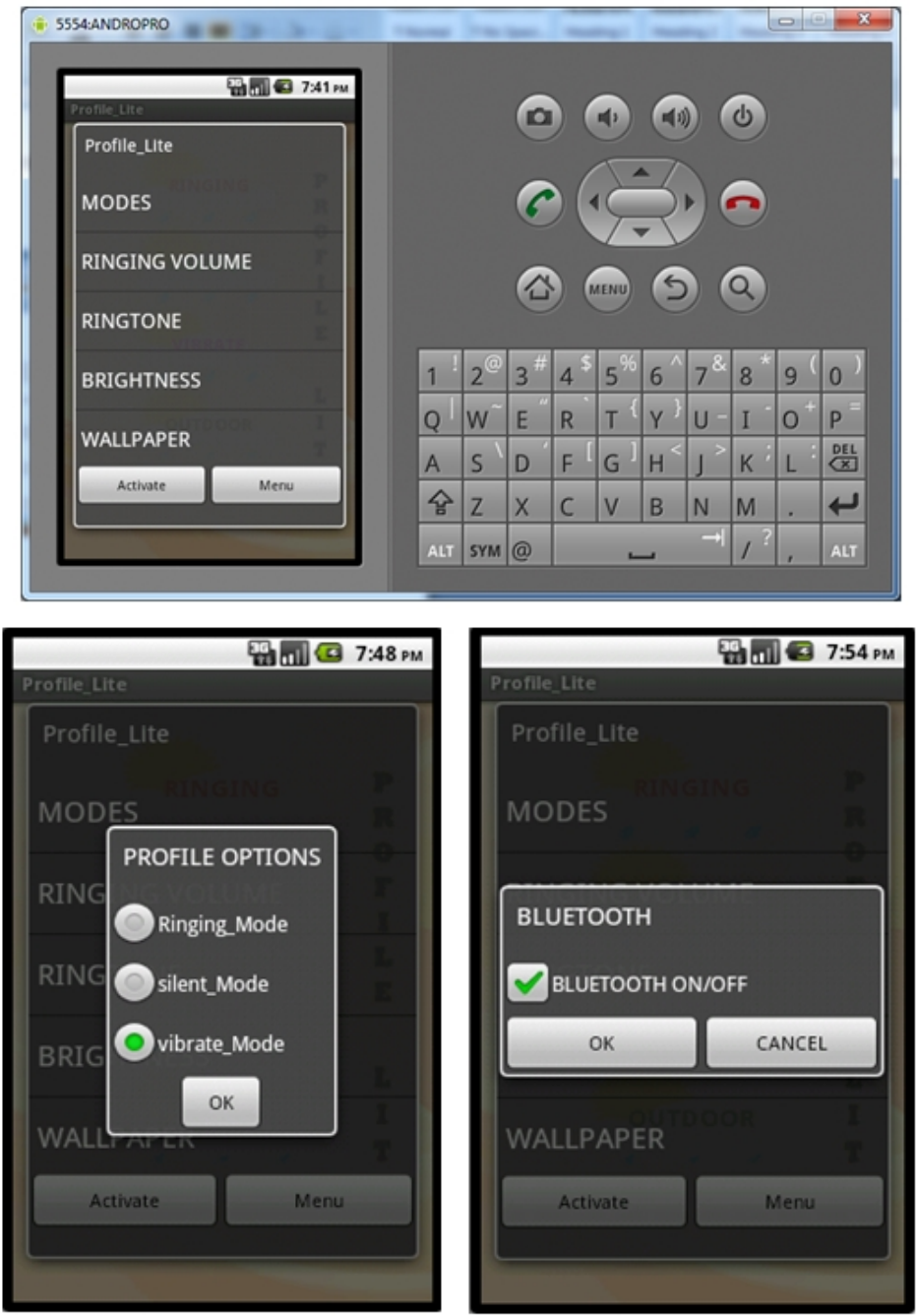

Figure 3: COSMOS Screenshot with Various Settings

Various screenshots of the COSMOS model are illustrated in Figure 3. It can be observed from Figure 3 that the interface provides various options to manoeuvre the settings. To confirm the efficiency of the proposed COSMOS model, various sessions of experiments were conducted on it. One of the parameters considered for the efficiency of the COSMOS model is by comparing the energy conservation made by making the optimal settings, which is illustrated in Table 1 and Figure 4.

Table 1. Battery Utilization Comparison

| Session ID | Mean Battery Hours (Normal) | Mean Battery Hours COSMOS |
|---|---|---|
| 1 | 17.2 | 18.3 |
| 2 | 16.5 | 17.8 |
| 3 | 18.6 | 19.8 |
| 4 | 19.1 | 20.3 |
| 5 | 15.4 | 17.4 |
| 6 | 16.8 | 18.1 |
| 7 | 18.1 | 19.6 |
| 8 | 19.3 | 20.8 |
| 9 | 18.5 | 19.9 |
| 10 | 19.2 | 19.8 |
| 11 | 17.5 | 18.9 |
| 12 | 16.5 | 17.9 |
| 13 | 16.8 | 19.7 |
| 14 | 18.8 | 20.3 |
| 15 | 19.4 | 21.1 |

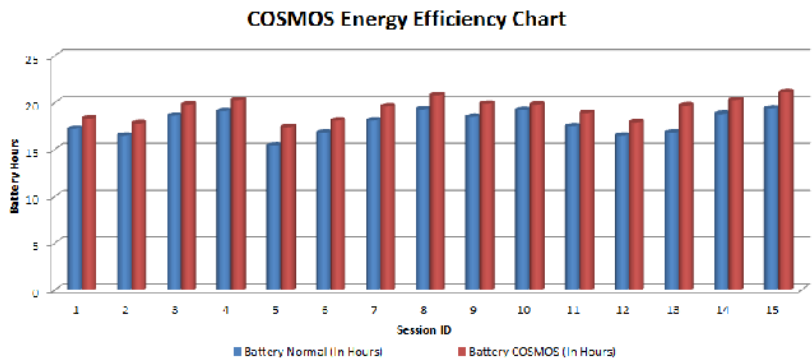

Figure 4. COSMOS Energy Efficiency Chart

It can be observed from Table 1 that, the mean of battery hours across all the sessions in the normal scenario is 17.84 whereas in the COSMOS the mean value is improved to 19.31 which indicate the energy utilized by the smartphone after the application of the proposed COSMOS model is efficient.

Although the settings of the smartphone are dynamically adjusted by the COSMOS model implicitly without interrupting the user, these settings need to be accorded by the user. The user's satisfaction level with respect to the automatic settings change is studied and the results are illustrated in Table 2 and Figure 5.

Table 2. COSMOS Settings Relevance Score

| Session ID | CRS | PRS | CIS |
| --- | --- | --- | --- |
| 1 | 72.4 | 20.1 | 7.5 |
| 2 | 80.1 | 10.3 | 9.6 |
| 3 | 82.1 | 12.8 | 5.1 |
| 4 | 69.5 | 20.5 | 10 |
| 5 | 70.5 | 15.6 | 13.9 |
| 6 | 75.6 | 14.5 | 9.9 |
| 7 | 80.5 | 10.5 | 9 |
| 8 | 88.5 | 6.5 | 5 |
| 9 | 84.2 | 10.5 | 5.3 |
| 10 | 81.5 | 9.5 | 9 |
| 11 | 83.4 | 10.5 | 6.1 |
| 12 | 88.1 | 4.5 | 7.4 |
| 13 | 80.2 | 5.9 | 13.9 |
| 14 | 83.5 | 6.5 | 10 |
| 15 | 79.5 | 6.5 | 14 |

In Table 2, CRS indicates Completely Relevant Settings, PRS indicates Partially Relevant Settings and CIS indicates Completely Irrelevant Settings. The comparison is charted out in Figure 5.

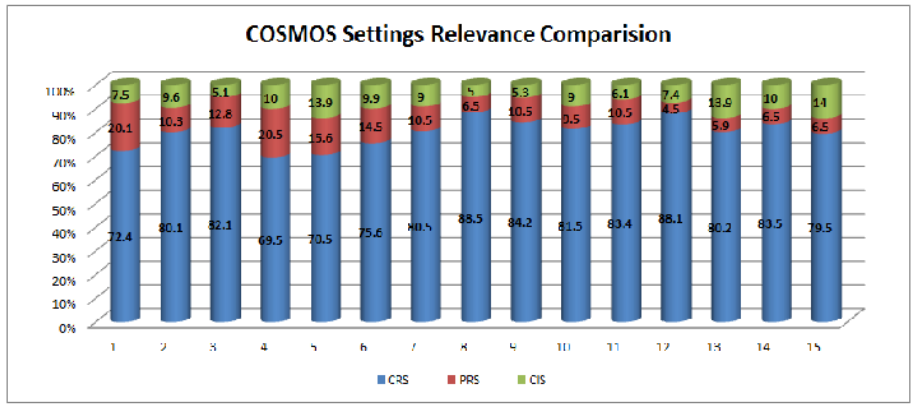

Figure 5. COSMOS Settings Relevance Comparison

It can be observed from data that 79.9% of settings changes made by the COSMOS model falls under the Completely Relevant category and 10.9% in partially relevant and 9.04% in completely irrelevant category. The cumulative of CRS and PRS is computed as 90.95% of settings changes made by the COSMOS model is accorded by the users as Relevant changes which confirms the efficiency of the COSMOS model.

## 5. CONCLUSION AND FUTURE DIRECTIONS

The proposed COSMOS model has been implemented in android operating system. It was tested in various devices from diverse manufacturers. The model can be implemented in other smartphone platforms like windows, apple, etc. The conclusions drawn from the proposed system are listed below:

- The COSMOS model facilitates dynamically configuring the smartphone devices based on the user context without interrupting the users.
- The settings chosen by the COSMOS model is tested for the accordance of the user's satisfaction level which was estimated at 90.95% confirming the relevancy and efficiency of the model.

The future directions for this research work are listed below:

- The COSMOS model can be further enriched by incorporating specific data mining algorithms in analysing various log data.
- The model shall be enhanced by widening its boundary by considering more settings and other services running in the device.
- The model shall be extended by considering parameters specific to Tablet devices in addition to the smartphones.

**Authors**

**Dr. K.S.Kuppusamy** is an Assistant Professor at Department of Computer Science, School of Engineering and Technology, Pondicherry University, Pondicherry, India. He has obtained his Ph.D in Computer Science and Engineering from Pondicherry Central University, and the Master degree in Computer Science and Information Technology from Madurai Kamaraj University. His research interest includes Web Search Engines, Semantic Web and Mobile Computing. He has made more than 20 peer reviewed international publications. He is in the Editorial board of three International Peer Reviewed Journals.

**Leena Mary Francis** was associated with Oracle as Software Engineer before she joined as Assistant Professor at Department of Computer Science, SS College, Pondicherry, India. She has obtained her Master's degree in Computer Applications from Pondicherry University. Her area of interest includes Web 2.0 and Information retrieval.

**Dr. G. Aghila** is a Professor at Department of Computer Science & Engineering, National Institute of Technology, Puducherry. She has got a total of 25 years of teaching experience. She has received her M.E (Computer Science and Engineering) and Ph.D. from Anna University, Chennai, India. She has published more than 75 research papers in peer reviewed, indexed international journals. She is currently a supervisor guiding 8 Ph.D. scholars. She was in receipt of Schrneiger award. She is an expert in ontology development. Her area of interest includes Intelligent Information Management, artificial intelligence, text mining and semantic web technologies.